\magnification=1000
\hsize=12.6 true cm
\vsize=19.5 true cm
\baselineskip=14 pt
\hoffset=1.5 true cm
\voffset=2.0 true cm
\def\ref{\smallskip\noindent\hangindent=1.0 true cm}
\vtop to 1.0 true cm {}
\centerline{\bf THE G DWARF METALLICITY DISTRIBUTION AND THE PROBLEM}
\centerline{\bf OF STELLAR LIFETIMES LOWER THAN THE DISK AGE$^*$}
\bigskip
\bigskip
\bigskip
\bigskip
\centerline{H. J. ROCHA-PINTO, W. J. MACIEL}
\bigskip
\centerline{\it Instituto Astron\^omico e Geof\'\i sico da USP}
\centerline{\it Av. Miguel Stefano 4200}
\centerline{\it 04301-904 S\~ao Paulo SP - Brazil}
\centerline{\tt helio@iagusp.usp.br, maciel@iagusp.usp.br}
\vskip 1.5 true cm
\centerline{\bf ABSTRACT}
\bigskip
\bigskip\noindent
We examine the possibility of the inclusion of objects having 
lifetimes lower than the disk age in the database of recent determinations
of the G dwarf metallicity distribution for the solar neighbourhood. 
As a preliminary step, stars with lifetimes potentially lower than the 
disk age are identified by means of a relation between $(b-y)$, or mass, 
and [Fe/H], from evolutionary models of stars at the zero-age
main sequence. We apply these results to the G dwarf samples of 
Rocha-Pinto and Maciel (1996) and Wyse and Gilmore (1995). 
We show that the majority of G dwarfs in these
samples can be regarded as long-lived objects for chemical evolution 
purposes, provided the disk age is greater than 12 Gyr. We also find
that the G dwarf problem is not an effect of the metallicity dependence
of the stellar lifetimes.
\vskip 3.0 true cm
\centerline{* Accepted for publication in the Astronomy \& Astrophysics.}

\vfill\eject
\noindent
\centerline{\bf 1. Introduction}
\medskip\noindent
The simple model (Schmidt 1963) was the first attempt to describe the 
abundances and the chemical evolution of the Galaxy. It explains fairly 
well the chemical evolution of closed systems such as the galactic bulge
(Pagel 1987), and is able to predict the correct order of magnitude of the 
radial gradients in the Galaxy (Maciel 1992), but it cannot account for 
the evolution of open systems like the disk, where gas inflow from external 
regions seems to have occurred. 

One of the best known problems with the simple model is the ``G dwarf
problem'', or its failure to explain the metallicity distribution of 
long-lived stars, as represented by the G dwarfs in the solar neighbourhood. 
The simple model predicts an excess of metal poor stars in the solar 
vicinity, contrary to what is observed (Pagel and 
Patchett 1975; Pagel 1989; Wyse and Gilmore 1995; Rocha-Pinto and Maciel 
1996; see Tinsley 1980 for a review).

The samples used to build the metallicity distribution of long-lived stars 
are based on G dwarfs, as these stars are thought to have lifetimes 
greater than or equal to the age of the disk, and can be considered as 
legitimate witnesses of the galactic chemical enrichment. However, there 
are some evidences indicating that the correspondence between the 
metallicity distribution of long-lived stars and that of the G dwarfs 
is not straightforward. Model calculations show that the lifetimes of the 
earliest G dwarfs could be slightly lower than 12 Gyr, which is a good estimate
of the disk age. Pagel and Patchett (1975) avoided this problem by 
eliminating all stars earlier than G2 from their sample. This simple 
procedure does not remove all stars with lifetimes lower than the disk age, 
as the spectral types depend both on the effective temperature and metal 
content. In fact, stars with the same MK classification may have very 
different masses (thus different lifetimes), due to variations in the
effective temperature $T_{\rm eff}$ and metallicity [Fe/H]. Moreover, the 
stellar chemical composition also affects its lifetime. Hence, a 
metallicity distribution based on a given range of spectral types could 
include stars that are not long-lived for chemical evolution purposes, 
even though classified as G dwarfs. 

Recently, Rocha-Pinto and Maciel (1996) presented a new metallicity 
distribution for 287 G dwarfs, using $uvby$ photometry and up-to-date 
parallaxes. The adopted sample includes all G dwarfs in the spectral 
range from G0 to G9 V, based on the argument that the hotter stars are not 
systematically more metal rich than the redder G dwarfs, as would be 
expected if they had shorter lifetimes. However, there is no clearcut
relation between the spectral types and lifetimes. The G dwarfs may
include stars within a given mass range belonging to different stellar
generations, thus showing different degrees of evolution on the main
sequence. As a result, a given region on the HR diagram may include
stars coming from several loci on the zero-age main sequence. According
to Nordstr\"om (1989), this effect produces an uncertainty of 15\% or
larger in the relation spectral type--mass. For a typical G5 V star,
with an estimated mass of 0.93 M$_\odot$ (Allen 1973), and a corresponding
lifetime of 13 Gyr (Bahcall \& Piran 1983),
this uncertainty would produce an equivalent uncertainty on the stellar
lifetime of the order of 6 Gyr. Thus, some G0 stars can be 
regarded as long-lived stars, while some G5 stars cannot. 

In the present paper, we propose an approximate method to identify stars 
with lifetimes lower than a given age, by means of a relation between 
the colour index $(b-y)$, or mass, and [Fe/H]. In section~2, we present 
the method for estimating the stellar lifetimes, and apply it to the 
samples of Rocha-Pinto \& Maciel (1996) and Wyse \& Gilmore (1995). 
In section~3 we discuss the effect of metallicity dependent
lifetimes on the G dwarf problem.
\bigskip
\bigskip
\centerline{\bf 2. Identification of short-lived stars}
\smallskip
\centerline{\it 2.1 The mass--metallicity relation}
\medskip\noindent
Stellar models provide lifetimes for stars of a given mass and
metallicity [Fe/H]. As the individual masses of the stars in the 
samples for which we want to apply our method are not generally
known, it is convenient to correlate the  stellar mass with 
some known property. In principle, if we have a temperature indicator for 
a star, its mass could be estimated by some astrophysical 
relations.

The method we present here uses the $(b-y)$ colour as a temperature 
indicator, that is, the same index adopted in the samples already mentioned. 
However, other indices could be used as well, and the same method could be 
applied to a mass--metallicity diagram. We preferred to use  a 
${\rm [Fe/H]} \times (b-y)$ diagram, since $(b-y)$ is the quantity 
directly observed. A similar procedure to exclude short-lived G dwarfs 
was independently developed by Wyse \& Gilmore (1995).

We have taken the lifetimes from a set of evolutionary models 
which cover a wide range in mass $M$, metallicity $Z$ and helium 
abundance $Y$ (VandenBerg 1985; VandenBerg \& Bell 1985; VandenBerg 
\& Laskarides 1987). 
These lifetimes were interpolated in order to derive a single relation 
between the mass and [Fe/H] for stars experiencing central hydrogen 
exhaustion $\tau_M$ years after their birth.  

The dependence of the lifetimes on the helium abundance was eliminated 
by means of the relation
$$Y=Y_p+{\Delta Y\over\Delta Z}Z,\eqno(1)$$
(Peimbert \& Torres-Peimbert 1974, 1976), where we used $Y_p=0.233$ 
(Chiappini \& Maciel 1994). The helium-to-metals enrichment
ratio $\Delta Y/\Delta Z$ was chosen in order to adjust the solar He 
abundance $Y_{\odot} = 0.27$ (VandenBerg 1983) at the solar metallicity 
$Z_{\odot}=0.0169$. It results that $\Delta Y/\Delta Z = 2.189$, which is 
close to the conservative value given in the recent literature (Peimbert 1996, 
Maeder 1992), and agrees very well with the observed main sequence 
width $\Delta M_v\approx 0.55$ (Fernandes et al. 1996). 
The metal abundance $Z$ was converted into the observed abundance  
[Fe/H] by the equation
$${\rm [Fe/H]} = \log\left[{Z\over 1-Y_p-(1 + 
    {{\Delta Y\over\Delta Z}})Z}\right] - 
    \log\left[{Z_\odot\over 1-Y_\odot-Z_\odot}\right].
    \eqno(2)$$

Figure 1 shows the expected [Fe/H] $\times\, M$ relation for 
stars having lifetimes of 9, 12 and 15 Gyr. It can be seen that,
for a given metallicity, the lifetimes decrease as the stellar
mass increase, corresponding to $\tau_M \ge 12$ Gyr for
$M = 1 M_\odot$ and [Fe/H] = 0, in good agreement with recent
calculations for the sun. On the other hand, for a given mass, the
lifetimes increase as the metallicity increases, which reflect the
fact that the opacities also increase (Bazan \& Mathews 1990). 
At very large metallicities ([Fe/H] $\ge$ +0.10 dex), the lifetimes 
tend to decrease, which is an effect of the increase of the helium 
abundances, according to the adopted evolutionary models.

The mass--metallicity relation can be fitted by a fourth-order polynomial
$$M = \sum_{i=0}^{i=4} \ a_i \ [{\rm Fe/H}]^i, \eqno(3)$$
and the coefficients $a_0, a_1, a_2, a_3$, and $a_4$ are given in 
table~1.
\medskip
\centerline{\it 2.2 The colour--metallicity diagram}

In order to transform the mass--metallicity relations into 
colour--metallicity [Fe/H]$\times (b-y)$ relations, the following steps 
have been taken: for each pair ($M$,[Fe/H]) we calculate the corresponding 
luminosity $L$ and radius $R$ at the zero-age main sequence (ZAMS), 
using the metallicity dependent 
mass--luminosity and mass--radius relations of Tout et al. (1996). 
These relations use essentially  the same $\Delta Y/\Delta Z$ 
ratio adopted here, which makes them particularly suitable. 

The luminosities and radii thus obtained are combined with the 
Stefan-Boltzmann law in order to derive the corresponding 
effective temperature, using $T_{{\rm eff}\,\odot} = 5780\,K$. 
Note that the results of the mass--luminosity and mass--radius
relations for a star having $M = 1 M_\odot$ and [Fe/H] = 0
are not expected to be exactly $L_\odot$ and $R_\odot$, as
they correspond to {\it zero-age} values. In fact, at the 
zero-age main sequence the sun had $L = 0.698\ L_\odot$ and 
$R = 0.888\ R_\odot$ according to the relations by Tout
et al. (1996), so that $\log T_{{\rm eff}\odot}({\rm ZAMS})
= 3.7486$, in excellent agreement with the result of the
solar model by VandenBerg (1985),
$\log T_{{\rm eff}\odot}({\rm ZAMS}) = 3.7480$.
\medskip
The colour index $(b-y)$ is related to the effective temperature
by the equation
$$(b-y) = -(1.94212 + 0.20792[{\rm Fe/H}] + 0.14239[{\rm Fe/H}]^2)
    \log T_{{\rm eff}}$$
$$ + (7.69792 + 0.81462[{\rm Fe/H}] + 0.54988[{\rm Fe/H}]^2),
\eqno(4)$$
which is derived from the theoretical colour--temperature relations
(VandenBerg \& Bell 1985) for dwarfs ($\log g \sim 4.50$).
This equation shows that for a given temperature, the colour $(b-y)$ gets 
bluer with decreasing [Fe/H], which is an effect of the line blanketing.

The derived colour-metallicity relations are shown in figure~2, and
these relations are fitted to fifth-order  polynomials
$$(b-y)= \sum_{i=0}^{i=5} \ b_i \ [{\rm Fe/H}]^i. \eqno(5)$$
The  coefficients $b_0, b_1, b_2, b_3, b_4$, and $b_5$ are also shown in 
table~1. It should be stressed that these relations are strictly valid
only for zero-age main sequence stars. Real stars show bluer colours
as they evolve on the main sequence, so that our results should be
considered as preliminary.

\midinsert
\par\noindent \bf Tabela 1.\rm Coefficients of Eqs. (3) and (5).
\medskip
{\halign{%
\hfil#\hfil&\quad\hfil#&\quad\hfil#&\quad\hfil#\cr
\noalign{\hrule\medskip}
$\tau_M$ & 9 Gyr & 12 Gyr & 15 Gyr \cr
\noalign{\medskip\hrule\medskip}
$a_0$  & $1.09226$   & $1.01776$   & $0.96091$    \cr
$a_1$  & $0.08527$   & $0.07701$   & $0.07528$    \cr
$a_2$  & $-0.18696$  & $-0.17460$  & $-0.16527$   \cr
$a_3$  &  $-0.15898$ & $-0.14492$  & $-0.14614$   \cr
$a_4$  & $-0.03650$  & $-0.03173$  & $-0.03444$   \cr
$b_0$  & $0.37422$   & $0.40739$   & $0.43930$    \cr
$b_1$  & $0.14870$   & $0.15263$   & $0.14007$    \cr
$b_2$  & $-0.09793$  & $-0.07868$  & $-0.05588$   \cr
$b_3$  & $-0.61373$  & $-0.61850$  & $-0.45216$   \cr
$b_4$  & $-0.58683$  & $-0.61440$  & $-0.40393$   \cr
$b_5$  & $-0.16743$  & $-0.18077$  & $-0.10171$   \cr
\noalign{\medskip\hrule}}}
\endinsert

We also plotted in figure 2 the 287 stars of the sample of Rocha-Pinto 
and Maciel (1996). Typical uncertainties in this diagram are $\delta{\rm [Fe/H]} 
\sim 0.16$ dex and $\delta(b-y) \sim 0.003$, as shown in the 
lower right corner of the figure. Stars to the right of a 
given curve have expected lifetimes longer than indicated by that curve. 
The dashed lines in this diagram correspond to the [Fe/H] $\times
(b-y)$ relations for stars of a given mass.

The obvious usefulness of fig. 2 is that it allows a 
straightforward identification of potentially long-lived stars. We can divide the 
diagram into four regions:
\smallskip
\par \noindent Region I: stars with $\tau_M<9$ Gyr.
\par\noindent Region II: stars with $9<\tau_M<12$ Gyr.
\par\noindent Region III: stars with $12<\tau_M<15$ Gyr.
\par\noindent Region IV: stars with $\tau_M>15$ Gyr.
\smallskip

The fraction of stars in each region is, respectively, 3\%, 15\%, 33\% 
and 49\%. The identification of the regions that are populated by 
long-lived stars depends on the assumed age of the disk. The most 
probable value for this age is around 12 Gyr, in any case not lower 
than 9 Gyr (Iben \& Laughlin 1989, and references therein; see also 
Cowan et al. 1991). {\it We conclude that stars 
belonging to regions III and IV are certainly long-lived, while stars 
in region I cannot be regarded as long-lived for  
chemical evolution purposes}. Therefore, approximately 82\% of the
stars in the sample by Rocha-Pinto and Maciel (1996) can be safely
considered as long-lived. Region II corresponds to the boundary of 
short-lived and long-lived stars. In view of the uncertainties in the
determination of the disk age, we cannot be sure whether or not 
they are representative of the chemical enrichment of the galactic disk. 
For a rigorous selection of long-lived stars, region II stars should be 
avoided.
 
We have also applied the present method to the sample of Wyse \& Gilmore 
(1995). The corresponding ${\rm [Fe/H]} \times (b-y)$ diagram for this 
sample is shown in fig. 3. In this plot, we have used different symbols 
to represent stars already excluded by Wyse \& Gilmore (empty squares), 
and those potentially long-lived (filled squares) for an adopted disk age of 
12 Gyr. In this case, the fraction of stars in regions I, II, III, and IV 
is  14\%, 13\%, 37\% and 36\%, respectively. Note that the percentage of 
their stars in region I is much larger that in our sample. This can be 
understood as Wyse \& Gilmore included some later F dwarfs in their sample. 

Wyse \& Gilmore (1995) excluded all stars with bluer colours than the 
turnoff colour at 12 Gyr, for a given metallicity. They have used the 
same set of stellar models (VandenBerg 1985; VandenBerg and Bell 1985; 
VandenBerg and Laskarides 1987), but do not mention how the helium 
dependence of the lifetimes was treated. In view of the 
similarities between our treatments, the good agreement between their
results and our 12 Gyr curve (see fig. 3) is not surprising. 

The results of figs. 2 and 3 are somewhat affected by the
evolution of the stars on the main sequence, as the colour-metallicity 
relations given by eq. (5) are only valid for ZAMS stars. On the other
hand, observational data refer to stars with varying degrees of evolution,
that is, their observed colours cannot be taken as zero-age colours.
To show this, we have considered the variations in $\log T_{{\rm eff}}$ 
between the ZAMS and the turnoff point given by Vandenberg (1985) and 
Vandenberg \& Laskarides (1987), for stellar masses $M = 0.8$ and 1.0 
$M_\odot$, $Y = 0.25$, and metallicities $Z = Z_\odot$, 0.03, 0.01, 
0.006, and 0.003. We converted the differences in the effective 
temperature into colour differences using eq. (4), and the derived 
results are shown in fig. 4a. It can be seen that the variation can
be large enough to move some stars born in the regions of long-lived
stars (regions III and IV) to regions I and II. 
This can be seen in fig. 4b for the case of the sun. In fig.1, the
solar position is slightly to the {\it right} of the 12 Gyr curve,
which would make the sun a long-lived star according to our previous
discussion. Taking the present sun colour $(b-y)_\odot = 0.405 \pm 
0.002$ (Saxner and Hammarb\"ack 1985), the position of the sun
changes to the {\it left} of the 12 Gyr curve, as can be seen in
figs. 2, 3, and 4b. This is basically due to the colour variation
on the main sequence. From fig. 4b, the sun was born on the 
1 $M_\odot$ curve, with approximately $(b-y)_\odot = 0.418$, that is,
in the region of the long-lived stars, in agreement with fig. 1. 
Due to main sequence evolution, it moves to the left, as shown in fig. 4b.
The arrow indicates the expected solar evolution from the variation
in the effective temperature according to the models by VandenBerg
(1985). As a general conclusion, we can see that
some stars located in the regions of short-lived stars could in fact have
lifetimes longer than 12 Gyr. Therefore, our results should be considered
as preliminary, and more accurate estimates are expected to be
possible as detailed calculations for evolved stars become available.
\bigskip
\bigskip
\centerline{\bf 3. Effects of  metallicity-dependent lifetimes} 
\medskip\noindent
Figure~5 shows the metallicity distribution of the sample by Rocha-Pinto 
\& Maciel (1996) in original form (solid line), and after 
eliminating stars with lifetimes potentially lower than 9 Gyr (dotted line) 
and 12 Gyr (dashed line). The total number of stars in each distribution
is 287, 278, and 236, respectively. As can be seen the elimination of 
the stars in regions I and II does not affect the shape of the metallicity 
distribution, except for a decrease in the number of stars
at the high metallicity end of the distribution.

These results indicate that, even if the disk age is about 9 to 10 Gyr, our 
sample is basically composed by long-lived stars. In fact, some 
investigations on the white dwarf luminosity function (Winget et al. 1987) 
and the evolution of the abundance ratios of the nucleochronometer pair 
Th/Nd (Butcher 1987; Morell et al. 1992) yield a disk age of 9-11 Gyr. 
On the other hand, although our distribution includes essentially
long-lived G dwarfs, it still cannot be considered as a complete
sample, since most old metal-poor G dwarfs with lifetimes of 12 Gyr or
greater are shifted towards the F dwarf range due to the evolution
on the main sequence, so that they are not selected as G dwarfs.

The legitimacy of the G-dwarf metallicity distribution as representative 
of the chemical enrichment history was first questioned by Biermann \& 
Biermann (1977). Taking the disk age as 15 Gyr, they concluded that only 
stars of type K0 and later can be considered as witnesses of the early 
phases of the Galaxy, and the customarily used G dwarfs are not old enough 
to represent these epochs. 

More recently, Bazan \& Mathews (1990) have suggested that part of 
the G dwarf problem could be an effect of the metallicity dependence 
of the stellar lifetimes. Their argument can be summarized as follows: for 
a given mass, the metal-poor stars should have shorter lifetimes than the 
metal-rich ones, as a result of lower opacities. Thus, as the older stars 
are likely to be preferentially the poorest, the number of expected metal-poor 
G dwarfs today should be smaller than the number of those stars ever born.

However, Meusinger \& Stecklum (1992) have pointed out that the results 
of Bazan \& Mathews could not be directly compared with the observational 
data, because Bazan \& Mathews define their theoretical G-dwarf sample 
by a mass range whereas the observed samples are defined by a spectral 
range and/or photometric criteria.

According to Bazan \& Mathews, the G dwarfs comprise stars between 
$0.79\,M_\odot$ and $1.09\,M_\odot$ (fig. 1, dot-dashed lines). This 
range is to be compared with the range predicted by Svechnikov \& 
Taidakova (1984), $0.90\,M_\odot$ -- $1.10\,M_\odot$ (fig.1, 
dashed lines). The upper mass limit is in very good agreement, but the 
value of the lower limit seems to be still uncertain. 

It is clear that if all stars born within the above mass ranges are 
G dwarfs, then we should expect that a large number of metal-poor G dwarfs 
will fall to the left of the 9 Gyr curve (cf. fig. 1), in the region of 
short-lived stars. As a result, part of the G dwarf problem could be 
originated by the mere absence of the old metal-poor G dwarfs from the 
data samples, because these stars have already died.

Note  that the majority of stars in our sample (which comprises only 
dwarfs with spectral type G) fall in the region limited by masses 
$1.1\,M_\odot$ and $0.8\,M_\odot$, in good agreement with the predicted 
G dwarf mass range (Bazan \& Mathews  1990), and to a less 
extent with the results of Svechnikov \& Taidakova (1984).

Besides the mass range, the G dwarfs are often defined by a colour 
range. According to Olsen (1984), G dwarfs have average
colours in the range $0.365\le(b-y)\le 0.470$,
which is in  good agreement with our results of fig. 2, in
the sense that 90\%  of all objects lie between these extremes.

In fact, the G dwarfs appear to lie in the very intersection between 
the mass and the colour ranges. Moreover, in spite of being well 
limited by the curves corresponding to their mass range, 
the observed G dwarf sample is better defined by the 
colour range. The dashed lines in fig. 2 cross regions of the diagram 
${\rm [Fe/H]} \times (b-y)$ which are photometrically associated to 
different spectral types. This indicates that stars of the same mass 
could have different spectral types. For example, the $0.9\,M_\odot$ 
curve shows that $0.9\,M_\odot$ stars would be F dwarfs 
[$(b-y) < 0.365$] at lower metallicities, G dwarfs for intermediate 
metallicities, and K dwarfs [$(b-y) > 0.470$] for [Fe/H] above solar.

From fig. 2 it can be seen that the region corresponding 
to the metal-poor short lived stars extends beyond the G dwarf range 
into the F dwarf range. Hence, {\it metal-poor G dwarfs will not have 
lifetimes lower than the disk age}, provided that this age does not 
exceed about 12 Gyr.

The short-lived G dwarfs are to be found mostly amongst the richest 
stars (see fig. 2), which is the opposite trend of that predicted 
by Bazan \& Mathews (1990). These G dwarfs are also massive objects, 
as shown by their location on  fig. 2. If they were also the youngest, 
probably all of these short-lived G dwarfs ever born would still be 
alive, as their expected lifetimes are between 9 and 12 Gyr. We 
conclude that the G dwarf problem is not an 
effect of the metallicity dependence of the stellar lifetimes.

The question remains as how to define selection criteria to assure the 
longevity and the completeness of the data samples used to derive the 
metallicity distribution. 

As we have shown, the selection of stars based on a mass range does not 
isolate the long-lived stars unless we take only stars with 
$M<0.9 M_\odot$ (see fig. 2). The same can be said about the selection 
by a colour range, unless stars redder than $(b-y)>0.430$ are considered. 
Selection by spectral type also does not isolate long-lived stars. The 
G dwarfs could be taken, in a first approximation, as  stars that live 
forever (that is, with $\tau_M\ge T_G$), as we have shown that 82\% of 
the G dwarfs are located in regions III and IV of the diagram 
${\rm [Fe/H]} \times (b-y)$. However, a metallicity distribution of K 
dwarfs could comprise only long-lived stars, as proposed by Biermann \& 
Biermann (1977). The extension of the G dwarf problem to later stars was 
examined in a few investigations based on small samples (Flynn \& Morell 
1997; Rana \& Basu 1990; Mould 1978), according to which the lack of 
metal-poor stars may also occur for redder dwarfs.

An alternative to this question, which we favour, is to make a selection 
by a lifetime range, for example taking only stars belonging to region~III 
in the ${\rm [Fe/H]} \times (b-y)$ diagram. In this case, not only 
G dwarfs must be taken in account, but also earlier K dwarfs and late 
F dwarfs, in order to check their position on the diagram. Once the
metallicity dependent mass-luminosity and mass-radius relations are
extended to evolved stars, it may be possible to solve the problems
associated with stellar evolution on the colour-metallicity diagram.
\bigskip
\bigskip
\centerline {\bf Acknowledgements}
\smallskip\noindent
We are indebted to Dr. P. Nissen and E.H. Olsen for some 
helpful comments on an earlier version of this paper. HJR-P thanks  
M\'arcio Catelan for some discussions and suggestions. This work was 
partially supported by CNPq and FAPESP.
\bigskip
\bigskip
\centerline{\bf References}
\medskip
\ref Allen, C.W.: 1973, Astrophysical quantities, Athlone
\ref Bahcall, J.N. and Piran, T.: 1983, ApJ 267, L77
\ref Bazan, G. and Mathews, G.J.: 1990, ApJ 354, 644
\ref Biermann, P. and Biermann, L.: 1977, A\&A 55, 63
\ref Butcher, H.R.: 1987, Nature 328, 127
\ref Chiappini, C. and Maciel, W.J.: 1994, A\&A 288, 921
\ref Cowan, J.J., Thielemann, F.-K and Truran,J.W.: 1991,ARA\&A 29,447
\ref Fernandes, J., Lebreton, Y. and Baglin, A.: 1996, A\&A 311, 127
\ref Flynn, C. and Morell, O.: 1997, MNRAS (submitted)
\ref Iben, I., Jr. and Laughlin, G.: 1989, ApJ 341, 430
\ref Maciel, W.J.: 1992, Ap{\&}SS 196, 23
\ref Maeder, A.: 1992, A\&A 264, 105
\ref Meusinger, H. and Stecklum, B.: 1992, A\&A 256, 415
\ref Morell, O., K\"allander, D. and Butcher, H.R.: 1992, A\&A 259, 543
\ref Mould, J.: 1978, ApJ 226, 923
\ref Nordstr\"om, B.: 1989, ApJ 341, 934
\ref Olsen, E.H.: 1984, A\&AS 57, 443
\ref Pagel, B.E.J.: 1987, in: The Galaxy, eds. G. Gilmore, B. Carswell,
     Reidel, Dordrecht, p. 341
\ref Pagel, B.E.J.: 1989, in: Evolutionary phenomena in galaxies, eds. J.E. 
     Beckman, B.E.J. Pagel, Cambridge University Press, Cambridge, p. 201
\ref Pagel, B.E.J. and Patchett, B.E.: 1975, MNRAS 172, 13
\ref Peimbert, M.: 1996, RMAA (in press)
\ref Peimbert, M. and Torres-Peimbert, S.: 1974, ApJ 193, 327
\ref Peimbert, M. and Torres-Peimbert, S.: 1976, ApJ 203, 581
\ref Rana, N.C. and Basu, S.: 1990, Ap{\&}SS 168, 317
\ref Rocha-Pinto, H.J. and Maciel, W.J.: 1996, MNRAS 279, 447
\ref Saxner, M. and Hammarb\"ack, G.: 1985, A\&A 151, 372
\ref Schmidt, M.: 1963, AJ 137, 758
\ref Svechnikov, M.A. and Taidakova, T.A.: 1984, SvA 28, 84
\ref Tinsley, B.M.: 1980, Fund. Cosm. Phys. 5, 287
\ref Tout, C.A., Pols, O.R., Eggleton, P.P. and Han, Z.: 1996, MNRAS 281, 257
\ref VandenBerg, D.A.: 1983, ApJS 51, 29
\ref VandenBerg, D.A.: 1985, ApJS 58, 711
\ref VandenBerg, D.A. and Bell, R.A.: 1985, ApJS 58, 561
\ref VandenBerg, D.A. and Laskarides, P.G.: 1987, ApJS 64, 103
\ref Winget, D.E., Hansen, C.J., Liebert, J., Van Horn, H.M., Fontaine, G., 
     Nather, R.E., Kepler, S.O. and Lamb, D.Q.: 1987, ApJ 315, L77
\ref Wyse, R.F.G. and Gilmore G.: 1995, AJ 110, 2771
\vfill\eject
\centerline{\bf Figure Captions}
\bigskip
\bigskip
\par\noindent
\bf Figure 1.\rm ${\rm [Fe/H]} \times M$ curves for stellar lifetimes
of 9, 12, and 15 Gyr. The vertical lines indicate the G-dwarf mass range 
according to Bazan and Mathews (1990; dot-dashed lines) and
Svechnikov and Taidakova (1984; dashed lines).

\par\noindent
\bf Figure 2.\rm ${\rm [Fe/H]} \times (b-y)$ relation for stars with lifetimes of
9, 12, and 15 Gyr (solid lines). Also plotted are the 287 stars in the 
sample of Rocha-Pinto and Maciel (1996, filled squares). An average error 
bar is shown at the lower right corner. The dashed lines correspond to the 
[Fe/H] $\times$  $(b-y)$ relation for stars of a given mass.

\par\noindent
\bf Figure 3.\rm The same as figure 2 for the 128 stars in the sample of
Wyse and Gilmore (1995). Different symbols are used to represent 
potentially short-lived (empty squares) and long-lived stars (filled 
squares), according to the criterion designed by Wyse \& Gilmore.

\par\noindent
\bf Figure 4.\rm (a) The same as fig. 2, showing the estimated evolution
    effects on stars of 1.0 and 0.8 $M_\odot$ with $Y = 0.25$ and 
    initial metallicities $Z_\odot$, 0.03, 0.01, 0.006, 
    and 0.003, respectively. (b) An enlargement of fig. 4a showing
    the variation in the solar position from the zero-age main
    sequence.

\par\noindent
\bf Figure 5.\rm The original metallicity distribution for the sample of 
    Rocha-Pinto and Maciel (1996, solid line), and after removal of 
    stars with lifetimes lower than 9 Gyr (dotted line) and 12 Gyr 
    (dashed line).

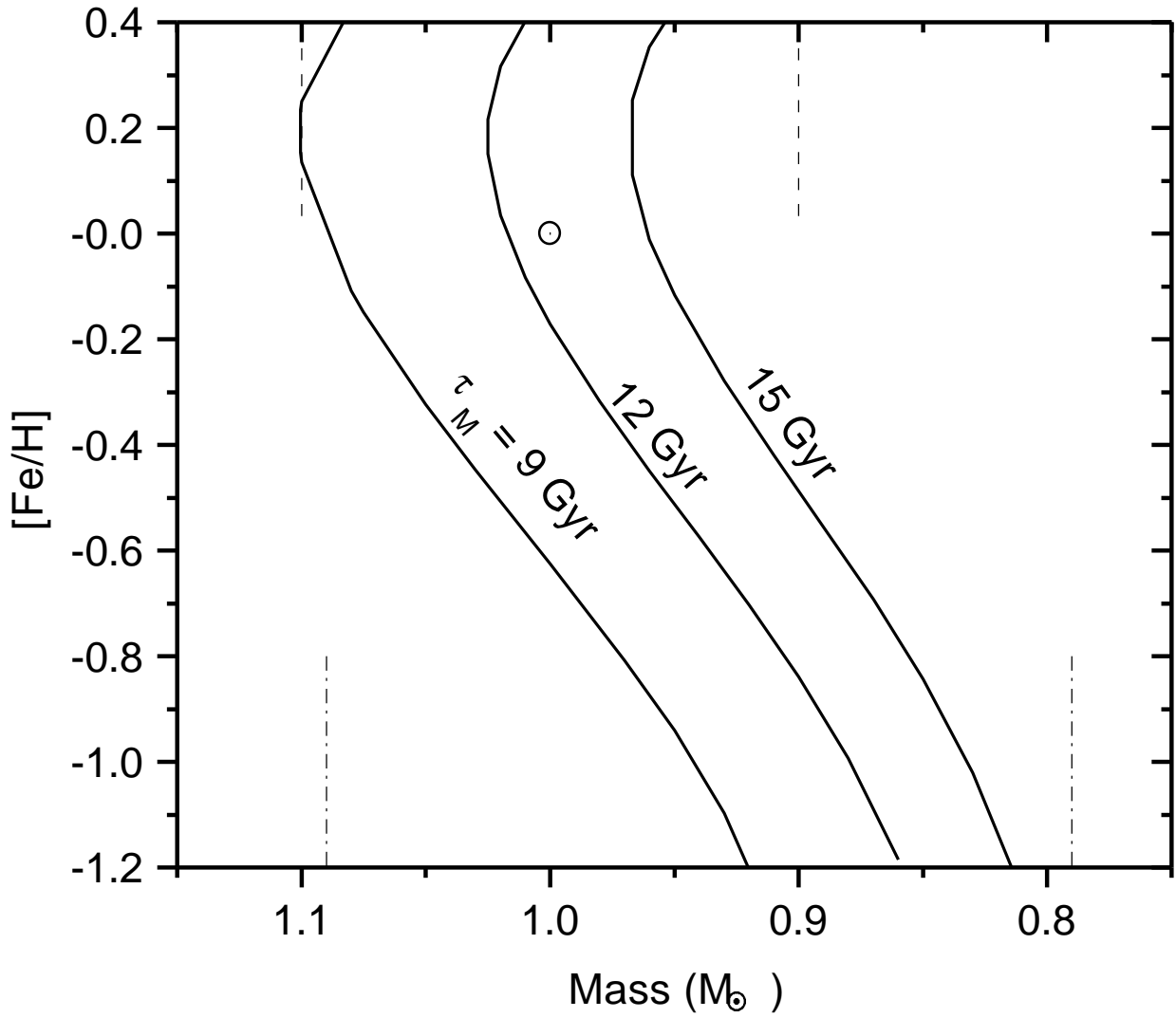
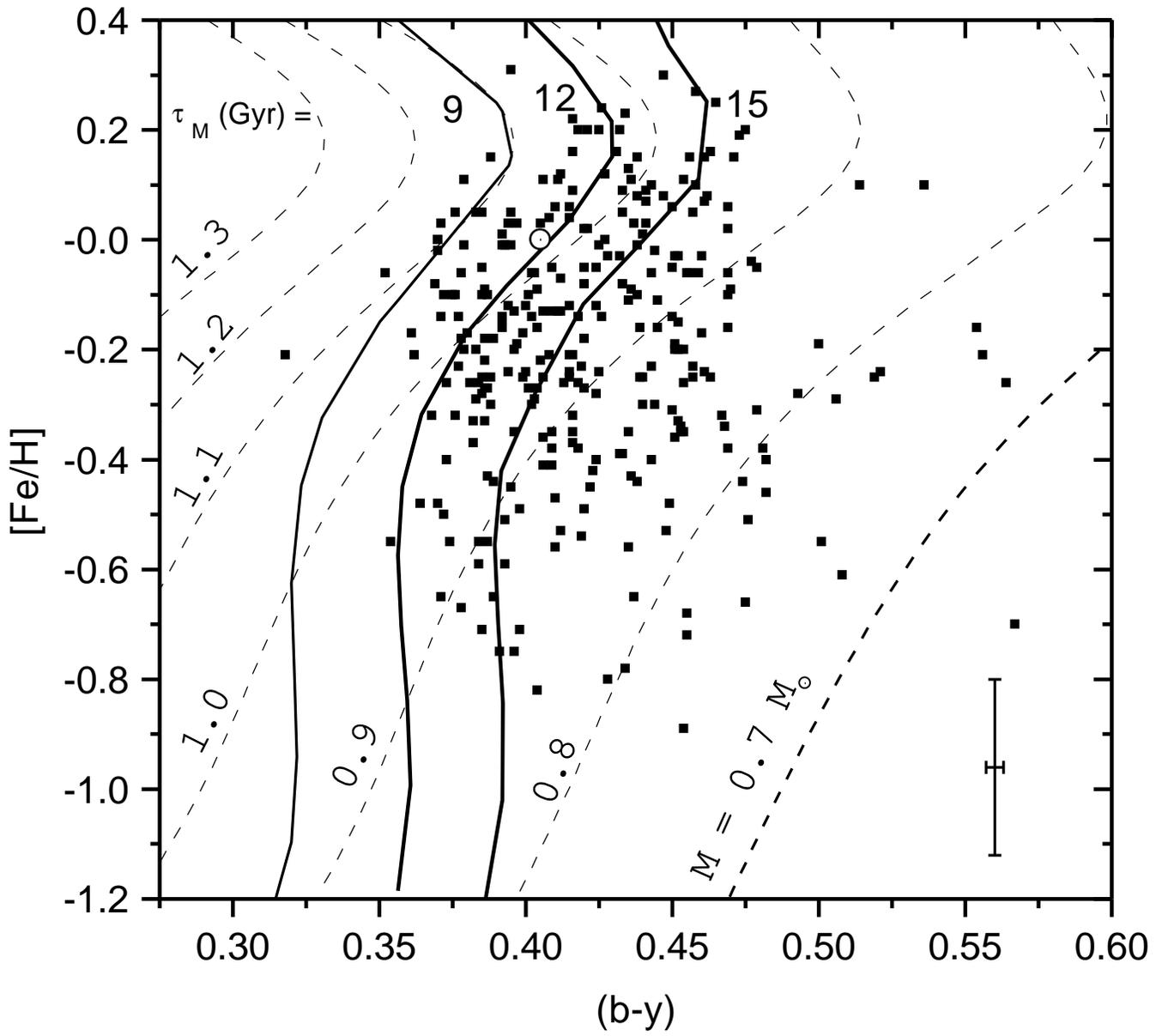
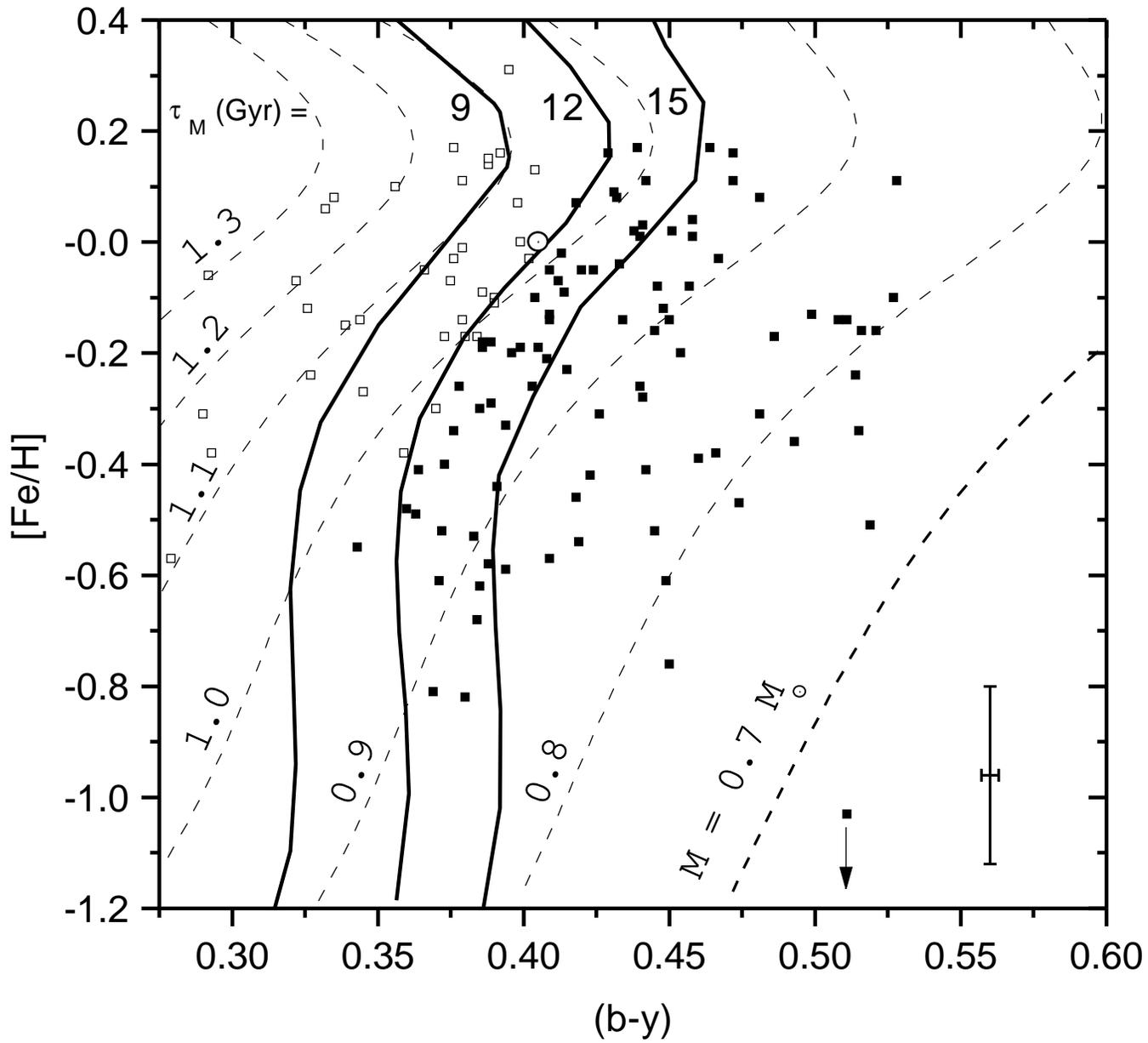
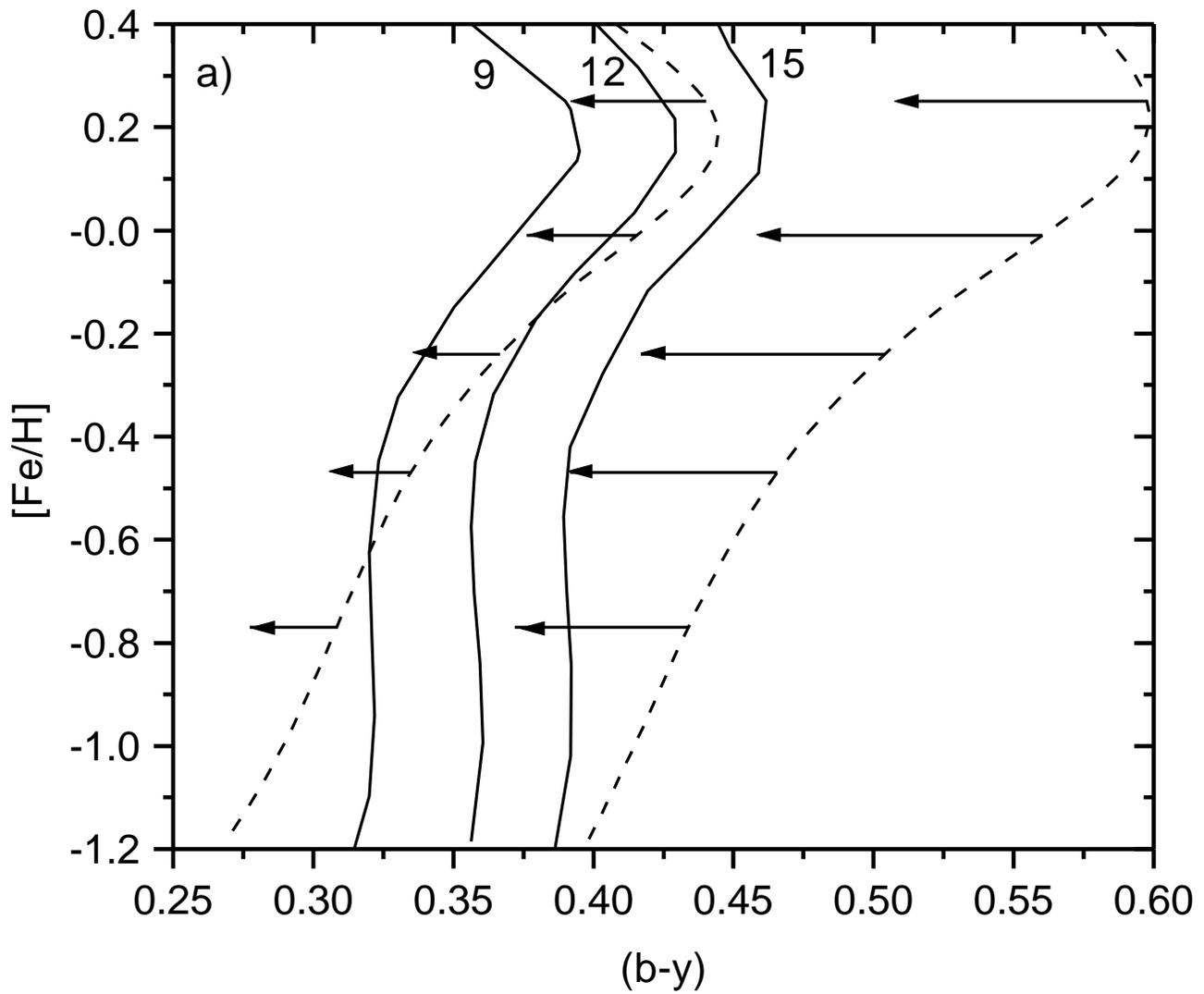
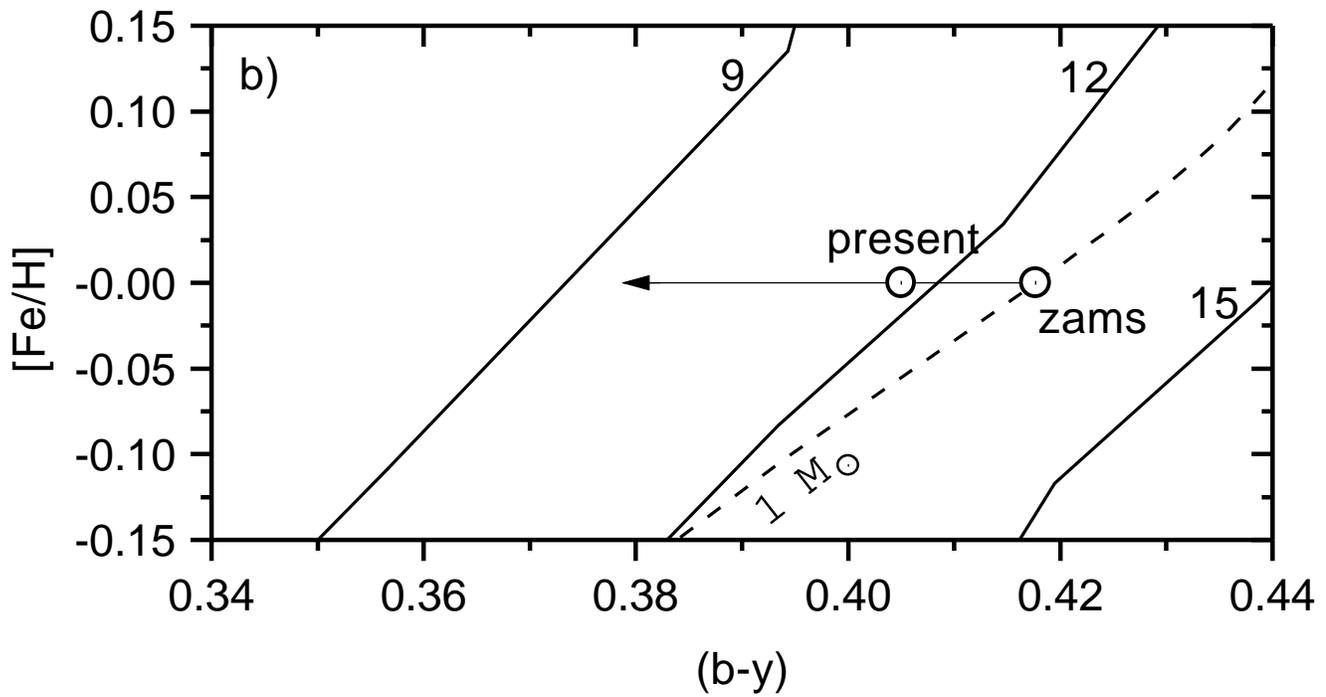
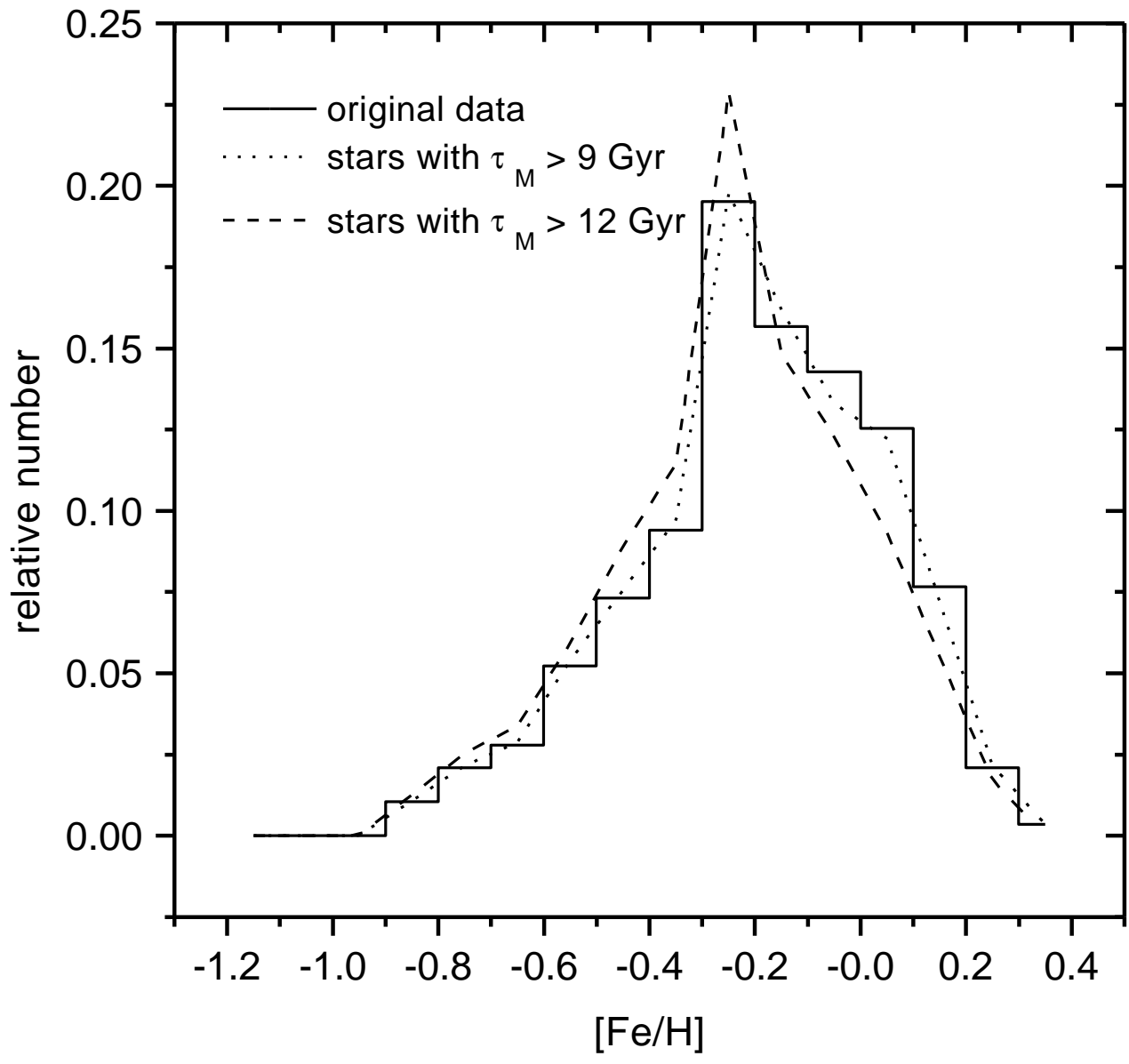
\bye